# Conceptual modelling for life sciences based on systemist foundations

Roman Lukyanenko[1], Veda C. Storey[2] and Oscar Pastor[3*]

*Correspondence:
opastor@dsic.upv.es

[1] McIntire School of Commerce, University of Virginia, Charlottesville, VA, USA
[2] J. Mack Robinson College of Business, Dept. of Computer Information Systems, Georgia State University, Atlanta, GA, USA
[3] PROS Research Center, VRAIN Research Institute, Universidad Politecnica de Valencia, Valencia, Spain

## Abstract

**Background:**  All aspects of our society, including the life sciences, need a mechanism for people working within them to represent the concepts they employ to carry out their research. For the information systems being designed and developed to support researchers and scientists in conducting their work, *conceptual models* of the relevant domains are usually designed as both blueprints for a system being developed and as a means of communication between the designer and developer. Most conceptual modelling concepts are generic in the sense that they are applied with the same understanding across many applications. Problems in the life sciences, however, are especially complex and important, because they deal with humans, their well-being, and their interactions with the environment as well as other organisms.

**Results:**  This work proposes a "systemist" perspective for creating a conceptual model of a life scientist's problem. We introduce the notion of a system and then show how it can be applied to the development of an information system for handling genomic-related information. We extend our discussion to show how the proposed systemist perspective can support the modelling of precision medicine.

**Conclusion:**  This research recognizes challenges in life sciences research of how to model problems to better represent the connections between physical and digital worlds. We propose a new notation that explicitly incorporates systemist thinking, as well as the components of systems based on recent ontological foundations. The new notation captures important semantics in the domain of life sciences. It may be used to facilitate understanding, communication and problem-solving more broadly. We also provide a precise, sound, ontologically supported characterization of the term "system," as a basic construct for conceptual modelling in life sciences.

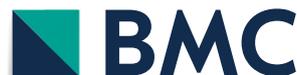

## Introduction

Life sciences are the branches of science that involve the scientific study of life itself. Understanding life has always been a major scientific objective and one that is increasingly important as we seek to understand the complex, global, and interconnected world in which we live. To understand life, we must represent or be able to model it [89, 113]. This requires the abstraction of concepts related to life sciences and their representation in a form that can be processed by an information system so that appropriate analysis can be performed.

As any live organism is complex, that is, made of interacting constituent components, all life forms are systems. To appreciate the entirety of the modelling of life, we need a precise, conceptual characterization of the "systems" that are involved in the working mechanisms of life. Life science is a natural context within which a well-grounded notion of system is crucial. Surprisingly, a sound, well-grounded conceptual characterization of "what" exactly a system is, does not appear to have been clearly stated in the literature, leaving an understanding of "system" open to interpretation. This lack of characterization of the system notion directly affects both the quality and correct use of life science information systems because different "types of systems" can be interpreted differently by the users involved in their creation and use.

The objective of this research is to refine and improve conceptual modelling activities by investigating a, thus far, overlooked concept of "system." We argue that, within the area of life science modeling, a holistic, *systemist* perspective is needed to achieve an integrated description of the types of problems scientists must analyze to manage the complexity of their work, all of which should be supported by a precise representation of the concept of "system." In this way, the notion of "system" becomes a basic conceptual modelling construct for information systems development, not only for the life sciences, but elsewhere as well.

Applying a systemic approach to life science would provide some immediate benefits in different, significant domains. For instance, it would allow to integrate different conceptual perspectives associated to the human genome understanding (the structural, DNA-based dimension, the transcriptional RNA-based one, the protein-based dimension, the functional pathways-based dimension, and the external manifestation (phenotype-based). These different perspectives are frequently managed in a separate way, in different data sources, with different data formats, making data integration a significant challenge. A systemist perspective would help to represent the relates aspects of data in a holistic way, helping to formulate more comprehensive solutions, treatments, and interventions.

Another interesting example of such a fruitful use of the systemist perspective could come from the semantic integration of the hierarchy of different locations that can be the source of genome data, what we could refer to as the "body location dimension" for particular genome data samples management. This hierarchical dimension goes from organs to tissues, cells or cellular components, each one with its own system representation that only through the use of a holistic systemist modeling perspective could be adequately managed.

A systemist perspective implies to need to use the concept of "system" as a basic conceptual modelling construct [100], analogous to how other constructs such as "entity",

"attribute", "role", "event," "agent" or "relationship" are used as basic modelling constructs in many diverse modelling domains and applications [24, 52, 81, 86, 151]. Despite the ubiquity of the notions related to systems in conceptual modeling and related areas of data management and software engineering, explicit modeling using systems as a construct is all but absent. Our paper addresses a call for explicit modeling of systems [100]. We do so in the life sciences domain but expect our approach to be generally applicable to other domains and use cases.

The explicit modeling of systems should provide value for different aspects of conceptual modelling. First, it can provide a way to represent the connection between physical and digital worlds in a sound and practical way. Second, it can facilitate semantic interoperability by integrating different systemist perspectives provided by different conceptual modelling languages that use a common notion of system while semantically referring to different kinds of systems. We refer here to the fact is that it is not sufficient to only model a system structure (the system components), without modeling system behavior and how the interaction between the system and its environment happens. Different models with different notations exist to face these different dimensions, while a unifying view is required to keep the global view that it is needed to manage the different parts that compose the system as a whole.

The contributions of the research are to: propose the notation needed to model a system holistically; and provide a systemist perspective for dealing with the complexity of precision medicine. To demonstrate the usefulness of the systemist perspective, we apply it to an information systems development problem involving an application to the human genome. The perspective is based on conceptual modelling so we: i) extend conceptual modelling expressiveness by introducing a well-grounded notion of system; and ii) show how this semantic extension can make life science information systems design and management more accurate and comprehensive.

In this paper we adopt a design science research (DSR) methodology [28, 85, 104, 138, 147]. Design Science research seeks to design useful artefact that provide solution within a particular context, while contributing generalizable design knowledge. An *artefact* is and object made by humans or machines with the intention to address a practical problem. Within the context of conceptual modeling and life sciences, our artifact is an innovative form of representation based on modeling systems. It is aimed at addressing a real-world problem of accurate and comprehensive description of facts in life sciences to improve our understanding of this domain and support more effective interventions.

Our research objective is related to both a knowledge question and a design problem. The knowledge question is: *How can we include the systemist perspective in conceptual modelling in order to understand how life science applications could benefit from it?* The design problem, with its associated artefact, is centered around proposing a concrete notation to model systemist expressiveness, showing initial examples of its practical application.

This paper proceeds as following. We first provide an overview of conceptual modelling. We then propose and illustrate a systemist approach to conceptual modelling and illustrate it with an application to the life sciences. The results are discussed with respect to broader implications that lead to significant questions that should be explored as future research.

## Systems in general discourse, sciences, and technology

### The notion of system in general discourse, sciences, and philosophy

This section reviews the usage of the system notion in general discourse and sciences broadly.

By system one generally understands something complex, with multiple parts, or some form of interrelationship and intricate organization. The modern term "system" is Greek (*systema*) or, possibly, Sanskrit (*samsthana*) in origin. In both languages these cognate words mean an organized whole, as well as have a sense of unity and "standing together" [65, 100]. Following this tradition, a conventional definition of a system emerged as a group of interacting or interrelated elements that form a unified whole.

In sciences, in addition to the common notion of system, another nuance is commonly added to the definition. Sciences understand systems not only as something which is complex, but also possessing emergent properties of behaviour [16]. In other words, systems are greater than the sums of their parts. This idea is rightly or wrongly attributed to Aristotle. In Metaphysics 8.6, Aristotle wrote ([11], p. 423):

In all things which have a plurality of parts, and which are not a total aggregate but a whole of some sort distinct from the parts, there is some cause; inasmuch as even in bodies sometimes contact is the cause of their unity, and sometimes viscosity or some other such quality.

This passage led to the common interpretation of Aristotle's conception of system as: "something over and above its parts, and not just the sum of them all" [56].

The idea of emergence gained prominence since Gestalt psychology in early 1900s. Psychologists such as Wertheimer [145, 146] and Koffka [90] suggested that humans perceive holistic images or patterns (or *gestalt*), rather than their individual components (a prior dominant view). In science broadly, the notion of system along with emergence continued to gain prominence.

As sciences probed deeper into the nature of reality, they nearly invariably uncovered complexity beneath simple appearances (e.g., discovering the components of atom, once thought to be indivisible). For example, a current debate in physics is whether *quarks* (presently considered elementary particles) are themselves made of particles [66]. Similarly, a heated debate in psychology (but also philosophy and theology) is whether a human is an *individual* (having single atomic identity) or a *dividual* (having multiple competing identities) [29, 82].

The scientific notion of systems gained further importance due to the rise of *chaos theory*. Chaos theory studies behaviour of complex, dynamic systems which are highly sensitive to initial conditions [1, 58, 150]. The theory describes how what looks like disorder (e.g., traffic jam, swarm of flies, human consciousness), emerges out of simple and deterministic laws which govern the components of these systems. Still, knowing the initial conditions as well as laws which govern the interaction among components makes it very difficult or at times, impossible to understand and predict what the result would like in some period of time [15, 54]. Effectively, many elements of existence, although obeying simple principles, take on a life of its own, having emergent properties.

Summarizing these developments, Stephen Hawking [83] proclaimed the twenty-first century is the "century of complexity." Indeed, not only in science, but in the society broadly, the pace of social, political and economic change appears to be accelerating [82, 132]. Complexity—the essence of systems—constantly gives rise to new products, technologies, concepts, and ideas.

### The notion of system in life sciences

Within the domain of life sciences especially, there is an important need to have system as a well-grounded construct. This construct is likely to be one of the most central constructs in the life sciences domain, along with other constructs, such as attributes and relationships. While conceptual modelling has been underrepresenting, ignoring, or not providing precise and unambiguous definition of a system, the domain of life sciences itself, had been founded around the notion of systems. The fields of physics, chemistry, biology, and more specialized disciplines, such as bioinformatics or biochemistry, all widely employ the notion of systems.

For example, such avant-garde thinkers in life sciences as Stephen Hawking, treats "system" as a fundamental constituent of reality. Hawking and Mlodinow [84] summarizing Feyman's theory of quantum mechanics, suggest that "a 'system', which could be a particle, a set of particles, or even the entire universe", may be predicted using Feyman's theory. This, they regard as the core strength of the theory. Hence, both in Feynman's formulation, as well as in the rendition by Hawking and Mlodinow [84], everything in the universe, from elementary particles to the entire universe, is a system, and furthermore, it is the objective of modern physics to predict the properties of these systems.

Likewise, in biology, the fundamental unit, the cell, is understood and modelled as a system. In chemistry and biochemistry, the most primitive entities, such as amino acids, the basic building blocks of life, are understood as systems [115]. Systemist perspective is foundational in ecology, which investigates the creation, evolution and destruction of ecosystems, such as lagoons or boreal forests [112]. Moving to higher order entities, such as human societies, sociologists have widely understood these to be systems [14, 30, 67, 94].

As a result of the centrality of systems to life sciences, a new interdisciplinary field of studies was born—general systems theory. The general systems theory was originally developed in the domain of life sciences by von Bertalanffy [19, 21]. In its original version, the theory suggested that, when conceptualizing life sciences, the notion of system is especially applicable. The theory understands systems as complex entities with interacting parts. Systems exhibit emergent behaviour (behaviour which cannot be deduced from the behaviour of the individual parts), feedback loops (with outputs of a system becoming its inputs), and homeostasis (the ability of the system to reach a certain level of stability).

It is further notable that toward his death, von Bertalanffy recognized the value of his ideas for the emerging fields of cybernetics and software engineering [57]. In his last essay, one month before death, von Bertalanffy [20] suggested that progress in life sciences and beyond can be accomplished with further developments in more accurate and complete modelling of systems, developments of sound engineering (including software engineering) practices based on systems thinking, and creation of new type of philosophy—"systems philosophy." This program was partaken both by information technology disciplines (including our focus: information systems and conceptual modelling) as well as philosophy. We review the status of these developments for modelling life science next.

### The information systems perspective

Early research in information technology development assumed that the objects of development are information systems; that is, complex objects, with interacting parts, that collect, process, store and manipulate information [51].

The notion of a "system" has been omnipresent in information systems (IS) research. As widely accepted in the information systems discipline, information technology artifacts (e.g., databases, application programmable interfaces, code routines, computer hardware) are components of larger systems, such as an organization or an ecosystem of other artifacts and its users, termed socio-technical systems [48, 102, 149]. Socio-technical systems are systems composed of technical components (processes, tasks, and technological infrastructure) and social actors (humans, their relationships and social structures) [27, 108, 114]. An information system, being a socio-technical system, binds together information technology, people, and organizations, produce joint outputs (e.g., information, furniture, raw materials, services) which they would be unable to produce separately.

Notions related to systems figure prominently in methods, theories, and approaches of information systems. An example of a systemic method is the *soft systems methodology*, a popular approach to business process modelling grounded in systemic notions [50]. For theoretical foundations, work systems theory argues that systems in organizations should be viewed as work systems [6, 7]. A work system is a system in which human participants *and/or* machines perform work (processes and activities) using information, technology, and other resources to produce specific product/services for internal and/or external customers [8], p. 5). The "work system" approach provides a natural unit of analysis for thinking about systems in organizations, emphasizing the need of having a well-defined systemist background intended to describe, analyse, design, and evaluate purposeful systems that perform work. Technologies, including information technology, should be viewed as components of work systems, rather than as systems on their own, unless dealing with fully automated systems.

To appreciate the systems perspective in information systems, consider an example of artificial intelligence in a hospital setting. Once a machine learning model is being used to treat cancer patients, it become part of a broader social system, such as a hospital. A hospital is a relevant example of a socio-technical work system, where a complex ecosystem connects the direct, physical, social interaction between patients, doctors, and the different hospital services (labs, hospital units) with the technical dimensions provided by the information systems that contain crucial electronic medical record (EMR, [116, 122] information (both conventional clinical data and genome-based data). Concrete biological artefacts can be the Variant Call Format (VCF) sample files (a logical artefact) obtained from a patient's biological samples (a physical artefact) that follow a genome sequencing process. The obtained information must be searched in other relevant artefacts of this ecosystem, as the genome variants data sources that contain the relevant information to be used to generate the final genomic clinical report.

### The system construct in conceptual modelling

As von Bertalanffy [20] noted, to understand and manage real-world systems, precise, complete and accurate modelling of these systems is needed. This is why is unfortunate that when it comes to modelling information systems with the help of conceptual modelling, systems modelling has been surprising neglected [100]. Still worse, system is a polysemic word and no agree-upon definition of system exists, including in information systems and conceptual modelling [51].

The notion of the system being part of conceptual modelling, has existed, arguably, from the early days of conceptual modelling theory and practice [57, 62, 130, 144]. As Mayr and Thalheim ([106], p. 2) note, "from the very beginning, conceptual modelling was propagated as a means to improve the design and implementation of whatsoever software system, especially with regard to a comprehensive and as clear as possible elicitation and analysis of system requirements" (p. 2).

The discussions of "system" in the context of conceptual modelling are present in earliest texts which laid the foundations of this new area of data management and software engineering. These included the consideration of the software being constructed with the aid of conceptual models as data systems, database systems or information systems (e.g., [2, 52, 55, 129] as well as of the domains which are represented in these systems as systems in their own right (e.g., [13], p. 10).

A system construct occurs indirectly in many conceptual modelling languages (for review, see [100]. First, the major constructs of traditional conceptual modelling grammars were assumed to be capable of representing systems. Conceptual modelling, fundamentally, involves modelling a domain using human concepts [76, 106, 110, 124, 134]. A core modelling construct, which emerged as early as the first conceptual modelling languages, is that of a *concept*, also known as a *category*, *entity type*, or *class* [98, 131]. For example, the UML Class Diagrams are structured around representing—the groupings of similar objects in the world which are modelled for the purposes of systems development [86]. Likewise, in the entity relationship model, entity types group related entity instances in the world [52]. In both cases, the objects in the world, entities, are complex,that is, systems. However, no specific distinction was made in these conceptual modelling languages between simple and structureless classes or entity types and complex entity types (systems).

Second, many conceptual modelling languages, such as UML class diagrams or extended entity-relationship diagrams, contain additional constructs, purportedly to represent notions related to systems. Hence, relationships constructs, such as "part of" in UML are systemic notions because they deal with complex objects (effectively systems) and their components. Systemic notions can be present in specialized conceptual modelling languages; notably, the Systems Modelling Language (SysML) [69], used for modelling engineering objects and processes. The repertoire of this language contains systemic notions, as the name of language itself implies. Still, it is difficult to find a precise, ontologically well-grounded definition of what precisely SysML means by a system.[1]

---

[1] By "ontologically well-grounded" we mean that the definition of a concept (system, in our case) has a sound conceptual background, ideally supported by a particular foundational ontology that captures precisely the conceptualization represented as an information artifact.

The references to "system" are rather generic and vague. They support the specification, analysis, design, verification, and validation of a broad range of systems, and systems-of-systems, without providing a precise conceptual characterization of what, exactly, comprises a system. As mentioned, the precise interpretation of what a system is, as well as the appropriate selected level of abstraction used for its representation, is left to the modeler. The challenge is that different users can have different intuitions about what system they are really describing, making it difficult to realize an accurate, shared communication about a system's description and its components. Nonetheless, the existence of this language underscores an appreciation of the need for having a more focused way to model systems.

Third, theoretical foundation of conceptual modelling, which uses the systemic notion, is *ontology* [73, 75–77]. Ontology is a branch of philosophy that studies what exists in reality, as well as what reality is. Existing ontologies which have been used in conceptual modelling also contain systemist notions. For example, a widely studied ontology in information systems, the Bunge Wand Weber (BWW) ontology, has a system construct [140]. In BWW, "a set of things is a system if, for any bi-partitioning of the set, coupling exist among things in the two subsets" [142], p. 222). It has been applied to prior work, including to the analysis and evaluation of constructs of conceptual modelling [22, 127, 139]. Another ontology with a systems construct is a realist ontology of digital objects and digitized systems has been proposed by Lukyanenko and Weber [101]. It aims to understand the nature of digital technologies and facilitate their design. This ontology, similar to BWW, contains the notions of systems, including system, level structure and emergent properties. Systems are considered, but not comprehensively modelled in other major ontologies, such as Descriptive Ontology for Linguistic and Cognitive Engineering (DOLCE) [25, 70] and Unified Foundational Ontology (UFO) [79, 80]. However, beyond an inclusion some systems-related constructs (e.g., part-of), existing ontologies used as references in conceptual modelling (including BWW), did not focus on systems specifically. The systemist ontological principles remain on the periphery of theoretical work in conceptual modelling.

Overall, systemic thinking, with its notions related to systems and the concept of a system, is deeply ingrained in the theory and practice of conceptual modelling. At the same time, the construct of the system itself is not explicitly present in any popular conceptual modelling approaches [43, 61, 64, 68, 101].

Shortcomings arise from the absence of the "system" construct from popular conceptual modelling languages. Considering the importance of system in sciences and daily life, Lukyanenko et al. [100] provided an analysis of the limitations of conceptual modelling due to the absence of the explicit systems constructs. In particular, they used a context of online citizen science[2] to show that conceptual modelling which does not involve explicit modelling of systems, may result in database structures which failed to adequately store the relevant information about the objects of interest and relationships among them. Absent explicit systems modelling, the quality of information captured in information systems may also be affected. Finally, they argued that without explicit systems modelling, it is very difficult to account for the emergent properties and behaviour of information systems, as well as a socio-technical work system which embed these technologies. Hence, they called for the development of systems conceptual modelling. Our paper addresses this call.

---

[2] Citizen science refers to various forms of involvement of the general public in scientific research. Citizens increasingly help scientists with data collection, analysis, and other research activities [23, 91, 93, 111]. Citizen science is also viewed as a kind of crowdsourcing—a way to organize work by outsourcing some of the tasks to online distributed crowds [46, 63, 97, 148].

## Foundations of modeling life sciences with systems

### The system construct in life sciences

The lack of explicit modelling of systems has general shortcomings in conceptual modelling, but it is especially pronounced in domain of life sciences, where system is a central and foundational concept. Our goal is, thus, to provide a precise, sound, ontologically supported characterization of the term "system," as a basic construct for conceptual modelling in life sciences.

Existing conceptual modelling languages, theoretical foundations of conceptual modelling, and methods for employing conceptual modelling languages in the development of conceptual modelling diagrams, have all avoided representing systems directly, treating systems as either the overall socio-technical systems in which a technology solution is being implemented or as entities which can be modelled using existing conceptual modelling languages. Even a specialized system modelling language, SysML [69], which put the notion of system at its core, lacked a precise and theoretically grounded treatment of the system.

Natural systems of special interest in our work are living organisms. Describing them in complete detail would, of course, be impossible mission. Therefore, a working context of interest, as well as an adequate level of abstraction must be selected in order to make it possible to obtain a precise conceptual characterization of the domain that is well-aligned with the purpose of the system description. For example, it is not as effective to merely identify the genome variants of a patient sample, as it is also to connect these variants with existing clinical knowledge, with specific diseases symptoms, and with the most convenient treatments. Connecting all of the relevant information that may appear in different formats, even analysed at different levels of abstraction or with a different granularity, requires a sound conceptual modelling effort.

A conceptual characterization of existing relationships between system/subsystem connection can help to identify relevant semantic connections among different system views (most likely represented in different conceptual models). If we assume a reductionism perspective, it may not always be possible to apply it. For example, living organisms might have subsystem decomposition that is not clear, or a reductionism assumption might fail. Even if we accept the fact that such an accurate model decomposition representation is not always possible, our work emphasizes the relevance of performing the conceptual task when different system representations exist. Furthermore, when connecting them as a whole, a systemist picture can be generated that provides a more clear and precise system description.

### The world as systems

Since von Bertalanffy [19, 20], there have been several notable attempts to provide a theoretical foundation for understanding systems (e.g., [3, 4, 14, 49, 94]. This systemist thinking in modern sciences has been captured, among others, by the philosopher-physicist Mario Bunge [31, 32, 34, 36, 37, 40]. The world, according to Bunge ([37], p. 25) is "made up of interconnected systems." We use Bunge as our foundation for system-based modelling in life sciences.

The ideas of Bunge are important for multiple reasons. First, they encapsulate the recent advances in life sciences, including the fundamental debates around the notion of system. Being a scientist himself, Bunge contributed to these debates, publishing in physics, chemistry, and biology outlets. Second, Bunge has been widely accepted as a reference for certain kind of conceptual modelling [77, 141, 143]. Hence, his work has already been accepted as relevant and useful for the development of conceptual modelling. Much of Bunge's influence has been via the BWW ontology, which has made substantial contributions to the theory and practice of information systems development and conceptual modelling. Lukyanenko et al. [99] further propose a new ontology, called the *Bunge Systemist Ontology (BSO),* based on the writings of Bunge later in his life (e.g., [36, 37, 40, 41], which is much more grounded in the systemist thinking than his earlier works (e.g., [31, 33, 34].

The *Bunge Systemist Ontology (BSO)* is based on Bunge's [37, 40] ontological claim that every *thing* is likely a *system.* This is an essential claim, and ontological primitive of BSO. According to Bunge, and BSO, the *world is made of systems*. Lukyanenko et al. [99] provide three explanations of this original ontological position.

First, *the notion* of a system allowed Bunge to reason about constituents of reality, which would be difficult to call *things*. For example, photon`s wave-particle duality or fields are better called systems. As Bunge ([40], p. 174) explains "the word 'system' is more neutral than 'thing', which, in most cases, denotes a system endowed with mass and perhaps tactually perceptible,we find it natural to speak of a force or field as a system, but we would be reluctant to call it a thing" (p.174).

Second, Bunge argued that *there are no simple, structureless entities*. Bunge observed that the history of science suggests that things, once thought to be irreducible and fundamentally simple (e.g., atom), have later proven to be complex. Bunge [40] concluded "*in tune with a growing suspicion in all scientific quarters*—that there are no simple, structureless entities" (p. 174, emphasis added). Indeed, this is exactly the conclusion reached by the pioneers of the quantum theory. Recall that, for Hawking and Mlodinow [84], everything in the universe, from elementary particles to the entire universe, is a system. Still this idea remains controversial in sciences, trailing behind the general controversial standing over quantum mechanics.

Thus, the idea that "there are no simple, structureless entities" is both an ontological, and a normative belief: "[t]his is a programmatic hypothesis found fertile in the past, because it has stimulated the search for complexities hidden under simple appearances" [40], p. 174). The complexity of a system can be hidden from our human perception capabilities,however, we assume it is always there. We do not take a pure reductionist approach, but, rather, argue that making it explicit (when possible) can lead to much stronger, precise, and well conceptually grounded system's descriptions.

Third, Bunge asserted that systemism provides an accurate approach for describing reality. It holds numerous advantages, in that it benefits from the advantages of other positions, without incurring their limitations [5, 36]. Specifically, systemism is a middle ground between individualism (which under-represents internal structures of a system, its relationship with the outer environment, its levels of composition and emergence) and holism (which is not interested in the components and specificity of subsystems).

Bunge offers a definition of a system as a "complex object every part or component of which is connected with other parts of the same object in such a manner that the whole possesses some features that its components lack—that is, emergent properties" [35], p. 20). We adopt the same definition in this paper.

Just about anything is a system in the domain of life sciences. Examples include: humans, other animals, viruses, organs, cells, DNA, amino acids, atoms, molecules, and more. This makes the explicit modelling of systems highly applicable to the life science domain.

A given system has the properties of its subsystems, as well as its own, which are the emergent properties. On methodological grounds, Bunge distinguishes two kinds of system: *conceptual* and *concrete* [35], p. 270). A *conceptual* (or formal) system is a system in which all of the components are conceptual (e.g., propositions, classifications, and theories). These systems reside the minds of *concrete* thinking systems (e.g., humans). The *concrete* (or *material*) systems are those made of concrete components such as atoms, organisms, and societies, and may undergo change. Concrete systems change in the virtue of energy transfer.

Different kinds of energy transfer occur in concrete systems, such as mechanical, thermal, kinetic, potential, electric, magnetic, chemical (e.g., in [39]. Energy transfer leads to change in states of systems, as they acquire or lose their properties, producing what humans conveniently label as *events* and *processes*. In contrast to concrete systems, conceptual systems do not change since they, themselves, do not possess energy. However, energy transfer occurs within and between concrete systems (i.e., humans who are thinking and communicating about these conceptual systems).

Bunge suggested that, to represent a system, four elements need to be described—Composition, Environment, Structure and Mechanism of the system. These are referred to as the *CESM Model*. The *composition* of the system are its components; the *environment*, the external systems with which the system and its subsystems interact; and the *structure*, the relations among its components as well as among them and the environment [34], p. 4). Mechanism is defined as "characteristic processes, that make [the system] what it is and the peculiar ways it changes" [39], p. 126). To illustrate how to represent systems using CESM, Bunge provides an example of a traditional nuclear family—a type of a social system [39], p. 127):

> *Its components are the parents and the children; the relevant environment is the immediate physical environment, the neighborhood, and the workplace; the structure is made up of such biological and psychological bonds as love, sharing, and relations with others; and the mechanism consists essentially of domestic chores, marital encounters of various kinds, and child rearing. If the central mechanism breaks down, so does the system as a whole.*

Based on CESM, then, system's boundary are those components which directly interact with the environment, whereas components that only interact with other subsystems of its parent system are internal components. The membrane of a cell is its boundary component, whereas its cytoplasm is internal.

While we adopted Mario Bunge's conception of system for our analysis, as we have shown, it is broadly consistent with how systems are understood in the life sciences. These ideas also suggest that the existing conceptual modelling approaches to systems are inadequate. The different conceptual dimensions that the CESM model introduces need to have a sound modelling support, including strategies to conceptualize system statics properties (related to (C)omposition and (S)tructure), system dynamic properties (related to (M)echanism), and system interaction properties (related to how relevant system properties are perceived or identified).

As Bunge's works suggest, representing systems goes beyond merely representing the components of systems. Rather, it involves capturing the environment, structure and mechanism of a system, and the system`s boundary and internal components, as well as the types of energy being transferred between systems. None of these nuances can be depicted explicitly in present conceptual modelling languages, thus creating a rift between the capabilities of these conceptual modelling languages and the nature of the fundamental constituents in the life sciences domains; that is, systems. Next, we attempt to address this limitation by providing a notation capable of modelling systems more explicitly, specifically aimed at supporting life sciences.

## Systems modeling for life sciences

### Proposed systems representation and example

Following systemist thinking, modelling efforts require a systemic level of description, which is frequently accomplished by identifying the modeler's purpose. Based on that purpose, an explicit selection is needed to establish the relevant system granularity. This decision will delimit the core of the systemist model to be build. If it is a data-centered representation of an organization (the static dimension, centered around composition and structure), a data model (e.g., ER Model, UML Class Diagram) will be the nuclear part of the model. If it is a description of the processes of an organization (the dynamic dimension, associated with the mechanisms), the central model will be a process model that describes its relevant tasks and activities (e.g., a BPMN model, a use case diagram). If the representation affects a Human-computer interaction strategy (the interaction dimension, the environment of the system), an interaction model will be in charge of delimiting which interaction units, user interface components and dialogs will be used, such as a presentation model [107, 123] or a CTT diagram [121].

As a system can be (naturally) composed by different subsystems, a main systemist model must adequately represent that set of dimensions in a conceptually, consistent way. The proposed notation must facilitate how:

1. To represent the holistic systemic composition of the system under analysis (the system together with its components)
2. To associate, to each initial system component, the set of model descriptions that characterize its intended purpose.
3. To facilitate the incorporation of a CESM perspective, allowing the accurate integration of different model fragments (for instance, functional interaction patterns (dynamic systemic view related to Mechanisms) associated to their conceptual data model counterpart (structural view related to Composition and Structure).

According to BSO the world is made of systems. The notion of "system" becomes the key modelling artifact. Thus, we introduce the "System Composition Diagram" as a root, first-level conceptual schema where basic system components are represented in a UML Class Diagram-like style. Rectangles represent systems that are connected by using system associations depicted by a < < system > > relationship stereotype. Semantically speaking, the < < systems > > relationship stereotype must determine what (sub)systems modelling components are being connected by the relationship.

BSO suggests that systems can be seen as complex objects with parts or components that are interconnected. BSO also captures how different (complementary) modelling dimensions might be considered to capture differences among them. The proposed notation supports these aspects.

Each system component can be associated to a particular model dimension where details about the considered system perspective are specified. Several model descriptions can be required in order to adequately describe a system. Addressing this perspective in the proposed notation includes the introduction of a Second Composition Level, which can be considered as a kind of "explode action" of the first-order level, which must include the system components required to be specified in further detail in the system under analysis. This appears analogous to established multidimensional analytical techniques, such as drill-down, which is used in online analytical processing (OLAP) for data warehouses. However, its support in conceptual modelling and ontological foundations is limited. The systems modelling presented here can provide such support, because the levels in the "System Composition Diagram" can be viewed as conceptual models for multidimensional storage applications.

This "explode action" representation can be iterated as needed to provide an accurate system description. This supports an adequate representation of the CESM view, allowing us to conceptually link structural system constructs with fragment process modelling constructs. Additionally, at any level, systemic or emergent properties can be specified through the corresponding attributes relevant for the particular level.

For instance, a system "Coronavirus Pathogenesis Pathway" can be initially associated to a Pathways Analysis Diagram (PAD). Figure 1 shows a diagram from QIAGEN Ingenuity Pathway Analysis, where the functional description of the different molecular processes is represented. Its main purpose is to describe how coronaviruses affect key cellular pathways enhancing their replication and virulence. It is obviously a complex system, where functionality is supported by biological objects that actively participate in the entire process execution. If these procedural steps are related to specific biological objects, a Class Diagram (CD) that includes these relevant structural participants should be introduced, facilitating an explicit identification and an accurate representation of the concrete relationships between PAD data components, Their UML CD counterpart must be explicitly stated. Figure 2 shows an excerpt of a Conceptual Schema of the Human Genome from (CSHG [118]). The purpose of the CSHG is to identify the relevant concepts that characterize genomic knowledge. Different views capture the genome-oriented DNA structure, the transcriptome-oriented RNA generation, the proteome-oriented catalogue of synthesized proteins, and the functional connection that delimits the role of the different cellular components that participate in pathways. In this example, for the sake of simplicity, we select the pathway dimension of the CSHG, which focuses on identifying the actors that participate in the pathway's execution, from a purely structural perspective.

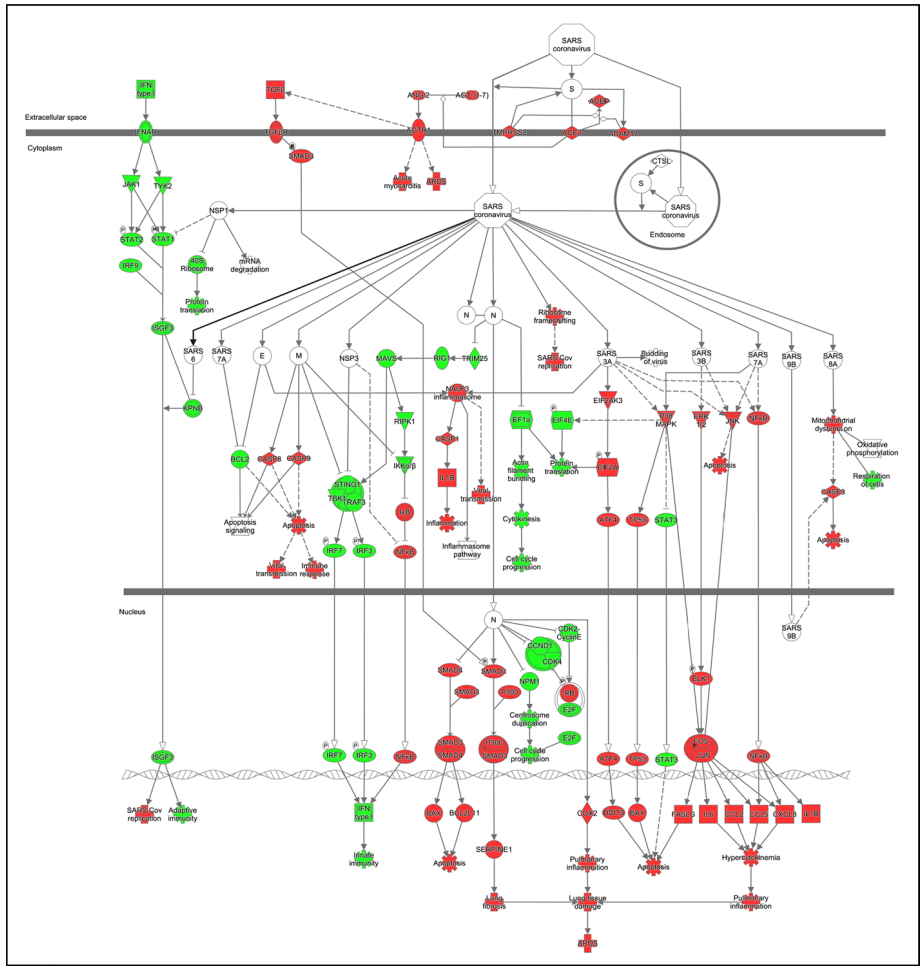

**Fig. 1** A Coronavirus Pathogenesis Pathway, from QIAGEN Ingenuity Pathway Analysis (IPA) (A full description of the associated functionality can be found in: https://www.researchgate.net/institution/QIAGEN/post/5fb69 62e4c5d8123c7042270_Coronavirus_Pathogenesis_Pathway_Poster)

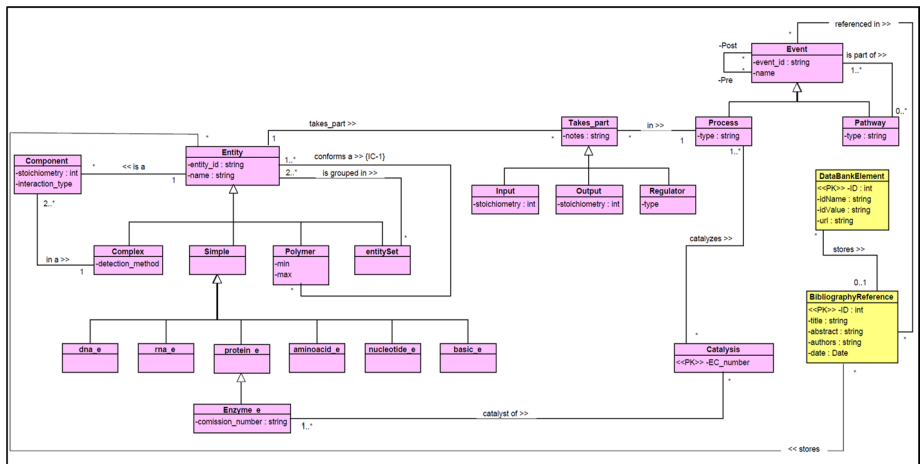

**Fig. 2** A pathway view of the CSHG. Pink boxes represent the classes that pertain to the Pathway view. Classes in yellow pertain to bibliography data

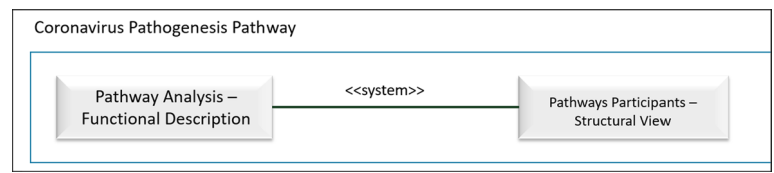

**Fig. 3** Systems configuration diagram representing the coronavirus

The systemist perspective (with its associated notation for practical modelling purposes), is intended to facilitate the integration of different dimensions of a given system. In this example, a functional pathway model (representing functionality, system's mechanisms) with a class diagram (representing participants' structure) must be adequately connected. This can be accomplished by creating a System Composition Diagram (for the coronavirus behaviour description) with two main system components: its functional pathogenesis pathway description; and the pathways view of the selected conceptual model of the human genome. Figure 3 shows the highest-level System Configuration Diagram that will model this system. The two relevant systems dimensions are represented by the corresponding rectangles:

1. Pathways Analysis Diagram (PAD), and
2. Pathways Structural View (UML CD (composition diagram) from the Conceptual Schema of the Human Genome).

The association between the two rectangles makes it possible to explicitly link structural components (e.g., proteins), with the functional role they play in the associated functional description (e.g., RNA viruses enter epithelial cells through specific docking proteins, such as ACE2 and TMPRSS2). This systemist, unified view is strictly required to well-understand the system under study. "Exploding" a system component will imply entering the associated systemist dimension description (in our example, either the PAD functional dimension or the CD structural dimension).

The Pathogenesis Pathway, thus, links two systemist dimensions: functional and structural. In terms of the CESM model, composition and mechanisms are conceptually integrated with the proposed representation. Additionally, the mapping between structural concepts from the UML CD and their functional PAD counterpart, as specified within the < < system > > stereotyped association, enables us to identify lacking connections. These connections can be highly relevant for facilitating a sound semantic interoperability between different systemist dimensions. For instance, if transcription regulators or transporter (which play a key role in the PAD model) do not have a structural counterpart in the CD, then, its conceptual representation needs to be adequately extended.

The purpose of introducing this system perspective is dependent on the specific system under analysis. In this example, the purpose is to improve the system description by making explicit and visible the set of semantic connections, what enables to better understand the system structure and behaviour. The motivation of the modelling task to be performed for formalizing the biological phenomenon under study is to facilitate data interoperability, integrating relevant data whose provenance can be diverse. This is only possible by providing a notation that makes the proposed systemist characterization feasible.

### Modelling with systems in life sciences

We can now show how to use our approach to model a system in the life sciences domain and demonstrate how a systemic approach is beneficial for doing so. Consider a typical healthcare domain, where the objective is to develop an electronic health record (EHR) application (see, [10, 42, 60]. Two different types of clinical information can be considered. On the one side, what we could refer to as "conventional" clinical data, that contains basic information associated to symptoms of a particular patient, given diagnosis, and selected treatments (e.g., [74]. For the sake of simplicity, we only consider these three main components of interest (symptoms, diagnosis and treatments), but more relevant components could be incorporated following the same conceptual model-based working procedure.

On the other side, we could consider the genomic roots that are associated to any particular health context. Two different systemist views emerge for these types of information: the more conventional electronic health record (EHR)-related one for the former, a genome-oriented one for the latter. The purpose of the system perspective consideration, in this case, is to make an accurate diagnosis and improving treatment efficacy by connecting two different (but complementary) systems, that become a whole system, a Person.

This scenario allows us to demonstrate the relationship between systems at different level, an essential dimension that is frequently ignored. Assume that the initial objective is to keep personal data of a Person (seen as a "patient" in clinical terms) together with a historical record of medical acts that include symptoms, diagnosis and treatments that the patient has had during her life.

If we consider a Person as a simple entity type, then a simple representation following a conventional data modelling strategy (including the subsequent relationships with other simple entities as Symptoms, Diagnosis and Treatments) is sufficient for a simple conceptual scenario where only such basic conceptualization is needed. However, if we consider a Person from the systemism perspective, a much richer "conceptual universe" emerges, conformed by a hierarchy of systems (as the Person genomic perspective) that would lead to a different level of description depending on the modeler's purpose.

We need to apply the system construct if we wish to conceptually characterize these two different perspectives and their connection: the Person from what we have called a conventional health-oriented perspective, and the structure of the patient's genome as a way to consider Symptoms, Diagnosis, and Treatments as an external representation and a consequence of much more complex behaviour that is the result of the internal genomic structure. Conceptually, the higher-level Person entity type would lead to a more complex, sophisticated system representation that will be dependent upon which one of the two contexts is specified, and how they are integrated (i.e., different phenotypes (associated with different diseases) will have different genomic reasons to justify a particular health problem).

As a Patient, a decomposed conceptual model could include a historical perspective describing the higher-level EHR associated with the patient's health history. This conceptual model (Person's EHR-oriented) should include new entity types that belong to the context being studied. However, a more in-depth conceptual characterization of the systemism dimension of the Person structure is needed if we want to connect the external manifestation of a health problem with its internal, genomic-based metabolic justification. This connection would be based on an associated conceptual model, focusing on a genomic-based system perspective (a subsystem of the Person system). Its relevant entity types would be those related to chromosomes, genes, variants, proteins, etc., all of which should be adequately connected to their functional environment (e.g., the external manifestation of a health problem) in order to determine pathological situations with clinical implications. Modelling these genome-based structure leads to another conceptual model dimension (Person's Genome-oriented).

As noted, introducing the system notion allows to conceptually characterize and manage the different system dimensions of the problem, and their relationships. For instance, the external identification of symptoms (a dimension in the Person's EHR view) can be seen as the tip of the iceberg, which shows us the manifestation of a health problem whose explanation is provided by its genomic roots. Identified genomic variants that are associated to a particular disease (at the Genome dimension) will be linked to symptoms that a patient shows (at the EHR dimension). The systemist perspective facilitates to model and represent the existing, conceptual connection, making visible at the "Person" root level, all of the relevant information that the subsystems capture.

Figure 4 presents an overview of a System Composition Diagram (SCD) for the discussed scenario, involving person and two subsystems. For the sake of simplicity, we represent the root system ("Person"), and its two subsystems ("Person EHR" and "Person Genome"). At the root level, a "Person" system includes two subsytems: the "PersonAsEHR" system and the "PersonAsGenome" system, together with a system association that connects them. Exploding (semantically speaking) the two systemic components leads to the associated UML CD, the conceptual model that represents for each subsystem relevant classes and relationships between classes. We use the UML CD standard notation, where rectangles refer to classes, lines connecting classes represent class associations, and association's minimum and maximum cardinalities are specified besides or under the associated classes using ''0', '1' or '*' -for many-. The system association is materialized through a class association between the "Diagnosis" class of the "PersonAsEHR" system, and the "Disease" class that represents a phenotype of interest in the "PersonAsGenome" system.

This initial systemism perspective can be extended if additional relevant information must be considered. In this basic version, treatments (in the Person-EHR system) are not connected with a different genetic background (from the Person-Genome system). The introduction of this concept in the System Composition Diagram would imply to the need to extend the < < system > > stereotype that connects the two subsystems with an additional, explicit association between Treatment (the Treatment class) and Variants (the Variants class). This association would enable a characterization when a treatment has different effects for persons, depending upon the particular genome variations of a person.

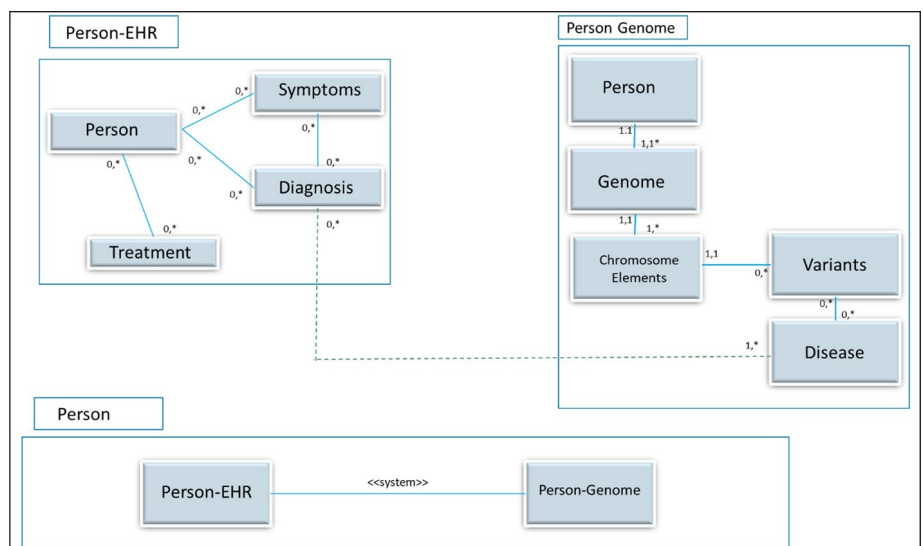

**Fig.4** A system composition diagram corresponding to a healthcare setting application

More "explode" levels could be explored; for example, the affected body parts (organs. tissues, cells, cellular components…) that conform to a subsequent system level. The genome sequence comes from a sample that is related to either a tissue or other relevant lower-level biological components. For instance, clinical focus can be put either on the hearth (as an organ), or on the epicardium, myocardium, epicardium (as tissues), or on the cardiomyocyte (as cell), or on the tubulles, sarcollema, sarcoplasm (as cellular component), all of which are hierarchically connected. This location-based connection between the organ affected by a symptom, and the genomic analysis of samples with the same source, could provide more internal, relevant system levels. An additional novel aspect of the proposed approach is how emergent properties can be identified and represented based on the use of the systemism perspective. Emergent properties can be distinguished from global aggregate properties. By global aggregate property we refer to those properties that are merely sums of the values of the properties of the components. For example, a definitive diagnosis for a person in the model in Fig. 4 (Person-EHR view), can be seen as simply the effect of the set of variants of a patient genome sample that can be modelled as a derived emergent attribute of the Person system (the top-root level of the model in Fig. 4).

In contrast, psychological anxiety about a given diagnosis in a patient can be viewed as an emergent property of the same Patient system that it is not the additive consequence of primitive psychological states; but rather, a consequence of the joint effect of properties coming for the different subsystems that the Systems Composition Diagram models. This situation can be modelled by specifying a new attribute in the Person system. Indeed, each system of the Systems Composition Diagram may have its own set of emergent attributes, either derived (i.e., when a decomposition for a higher-level property is inferred from other simpler subsystems properties), or not.

It is the modeler who decides when a property should be modelled as a global aggregate property or an emergent property. Regardless, having the capability to select the correct conceptual characterization is crucial for making a correct data management policy feasible. This is the main contribution of this research.

This systemism perspective can facilitate adequate data management within the context of a new medical practice or the *Medicine of Precision*, where a symptom is not simply the external perception of a situation [12, 72]. It is also the biological result of a specific genomic alteration, that can be detected and identified even without the need for "visualizing" the external effect of the variant problem. This example can show how a BSO-based ontological description of a Person entity type could lead to the design of a hierarchy of related conceptual models. Hence, systems of models can mimic the systems in the real-world: going from the external, visible "Person" features to the patient's affected, to relevant organs and tissues and to sequenced genome and its associated metabolic-based behaviour. These models are connected by the systemist level structure (i.e., composition and emergence), thus providing the relevant system descriptions at the different levels of detail required by the different context of use, each of which is based on a sound ontological background.

## Discussion

The modelling of life sciences based on systemist foundations is expected to offer benefits for modelling, understanding and management of life sciences, including for development, management, and use of life sciences information systems.

### Systemism for life sciences

Life scientists invariably use information systems to support their research. The modelling of such systems is always a prerequisite for the development of such information technologies. This research has developed an extension to popular conceptual modelling techniques that include the notion of a "system." This is significant in the life sciences as the domain of sciences is made of complex objects, that is, systems.

To support the continued digitalization of life sciences, we proposed a new notation that explicitly incorporates systemism thinking, as well as the components of systems based on recent ontological foundations. The new notation captures important semantics in the domain of life sciences and may be used to develop information technologies in this domain, as well as to facilitate understanding, communication and problem-solving, more broadly.

Among other uses, the systems-based notations we introduced can facilitate the application in use of big data and big data-based approaches, such as machine learning in life domains. For example, in healthcare, the use of big data (i.e., data characterized by large volumes, velocity, and variety), have led to better diagnosis, treatment and increased operational efficiencies [44, 88, 109]. Artificial intelligence (AI)-based techniques, such as machine learning increasingly leverage the availability of data (including big data), to automate medical diagnoses and treatments. However, among the difficulties in using big data and machine learning is ensuring development teams fully understand the domains. This can be quite challenging when dealing with complex life science domains, resulting in numerous cases of biases and errors in AI-based medical applications [9, 133, 135, 137].

Recently, conceptual modelling have been suggested as one way to improve transparency of machine learning development process and help developers better understand domains [26, 95, 96, 103, 125]. The new conceptual modelling notations that are introduced in this paper, can be especially powerful to unlock the potential of big data and AI in healthcare by providing a conceptual characterization of what data is of interest that must be analysed. It can also offer a sound bases for continued improvement of conceptual modelling for machine learning techniques.

One significant aspect that requires further research work is how to incorporate the Environment CESM element in the proposed modelling notation. Modelling the environment should make it possible to distinguish between the external systems with which the system and its subsystems interact, and the internal systems that have a common compositional architecture. Our current proposal does not yet cover this dimension, so a new modelling element is needed.

### Managing complexity

Systems-based modelling can be especially valuable for coping with the vast complexity of entities in life sciences. Actual complexity is the an absolute or relative (to other systems, domains) number of components of systems along with the way in which these components are structured and interact with one another and with other systems [38, 92, 100]. In contrast, perceived complexity is human's impression and conceptualization of a system as being complex [92, 126]. The complexity challenge in life sciences is both actual and perceived, as vividly depicted in an anecdote by Albert Szent-Györgyi, a Nobel laureate for isolating vitamin C [19], p. 5):

> *[When I joined the Institute for Advanced Study in Princeton], I did this in the hope that by rubbing elbows with those great atomic physicists and mathematicians I would learn something about living matters. But as soon as I revealed that in any living system there are more than two electrons, the physicists would not speak to me. With all their computers they could not say what the third electron might do.*

Systems modelling proposed here can help cope with actual and perceived complexity by abstracting from specific systems attributes at different levels, as well as guiding the reader in understanding how system components interact and systems at different levels are structured.

Following Guizzardi et al. [78] and Villegas Niño [136], we first emphasize the three main methods normally used in the discipline of complexity management of large conceptual models: clustering, relevance and summarization. We provide a case for a clustering method that is focused on the ontological BSO-based notion of system. Using the BSO notion of *system*, a concrete cognitive justification is present. The systemic perspective of a System Composition Diagram facilitates the selection of the right modelling components, thus providing a clear criterion for classification, with the system notion being the basic conceptual building block.

We also propose the notion of system-centric conceptual model modularization. The components of the system are clearly represented in the multi-level conceptual structure. A root System Composition Diagram delimits the systemic components of interest that can be decomposed in as much additional exploded levels as required. Different, well-known, and widely used modelling approaches can be used to specify diverse systemic dimensions of interest (e.g., structural, functional, interaction-oriented) [53, 59, 77, 128].

Systems (identifying system components) and system associations (denoting how they are connected) are first needed to cover system composition and structure. The multi-level dimension of the system specification helps to integrate different modelling perspectives while exploding the level of systemic description. Each model fragment can represent a system dimension, with each one at the right level of the abstraction following the "system explode" methodological strategy.

The incorporation of *Mechanisms* of CESM can conceptually be handled in a natural way using functional model fragments that adequately connects to the compositional and structural systemic dimensions. It enriches and extends them with the specification of the processes that characterize what the system is and how it changes over time.

After a design template for representing systems has been proposed, domain experts must be able to design correct models and understand their contents. When developing information systems, complexity is a significant challenge [87]. Designers must assess whether: the additional design work that is introduced is clearly identified; an illusion of simplicity for the selected representation is retained; and an effective visual representation of the elements of the model is created. Recall that a main purpose of the notation purposed is to incorporate adequately the CESM model in the modelling efforts. Although this is a major undertaking, this paper introduces a concrete, initial proposal for including adequate complexity management mechanisms.

We also note that our notation conforms to a sound ontological foundation, and hence, is expected to be effective and reliable. A common problem for many modelling languages is the lack of a precise ontological semantics [78]. However, this is not the case here. The BSO provides the underlying ontological theory that we use to provide sound language support. Within the context of a well-grounded ontology-driven conceptual modelling (ODCM), our goal is to design a language that conforms to BSO and allows us to represent systems as clearly as possible. Examples of similar approaches can be found with OntoUML and the UFO foundational ontology [77], or with the OO-Method and the O3 foundational ontology [117, 120]. Our work adds to this list.

### Integrating different perspectives of genomic knowledge

When dealing with genome data, a rich set of data formats is needed. These represent the relevant signals that cover complex and diverse aspects of how to correctly represent DNA variations, how gene activity is expressed, or in what way structural rearrangement of DNA works. Different conceptual models are typically used to explain these complex structures and behaviours. Integrating different perspectives into a common conceptual model, per our formalized, would help to connect the various views and solutions into a coherent hole, corresponding to the natural systems, such as DNA, genes, proteins, and cells.

As a concrete example, we consider the set of works produced by the PROS research centre summarized by the Conceptual Schema of the Human Genome (CSGH)[119], where (conceptually) a top-down representation of the genome is provided. Having in mind elaborating a general understanding of the language of life [71], it focuses on the concepts that characterize the genome components, independent of concrete (but diverse) data formats that are used in common practice.

Another very relevant initiative is represented by the GeCo project [47] and its Genomic Conceptual Model (GCM) [17]. Taking a complementary point of view (focusing on data), it provides a data-oriented, bottom-up representation, targeting a high-level, abstract description of these formats, that emphasizes what data is captured, how the process of capture is performed, and how those different signals can be integrated. Further exploring this approach, important results have been obtained in the area of data integration and search systems for genomics researchers [45, 105].

The integration of these two different perspectives—that we refer to as "top-down" (from concepts to data) and "bottom-up" (from data to concepts)—would directly benefit from using a systemist approach as the one presented in this paper. This approach would facilitate the description of a unified conceptual model that integrates higher-level genome concepts and a data layer where these concepts are represented as data that are obtained from practice. The initial results of this promising scientific task are already in progress [18]. The use of the systemist modelling approach would enable us to approach this task in a more natural, clear and, subsequently, valuable and effective manner.

## Conclusion

Unravelling the secrets of human conditions and diseases is a major challenge in the domain of life sciences. To achieve it, a precise ontological characterization of the relevant components of complex biological processes is required. This research proposes a well-grounded ontological notion of *system* as a basic modelling unit to effectively support the modelling of biological processes. The system construct enables the models to represent complexity inherent in all systems, including their systemic components, their interactions and the interaction between components and their environment, as well as the mechanisms by which the systems change.

A concrete representation strategy to model "systems" is proposed. The selected notation includes a Systems Composition Diagram that delimits a root, first-level conceptual schema from which different (sub)system components can be introduced and interconnected. These explicit connections are expressed through a < < system > > stereotype. Different systemist dimensions can be incorporated into the model by using the most adequate model representation, which is the Systems Composition Diagram artifact that retains semantic consistency. Exploding higher level classes using the appropriate modelling technique enables the exploration of conceptual hierarchies that can be adapted to the selected abstraction level.

This research has been illustrated with examples based on the human genome and viral diseases to illustrate how this modelling strategy can be applied in practice. Although more research is required to thoroughly evaluate our proposed notations and extend them, if necessary, the incorporation of the proposed systemist perspective to the design and development of life sciences information systems is already possible. Additionally, the proposed approach allows us to infer new, valuable information from the systemist relationships that are captured. A holistic systemist description is facilitated by interconnecting the different relevant system dimensions, which is something that, all too frequently, is not present in life science-based practical information systems.

Modelling life systems based on systemist foundations promises to improve the current practices in data representation and data management, making it also feasible to gain richer semantic interoperability among different components of information systems designed with the help of our proposed notation.

### Abbreviations

| | |
|---|---|
| CM | Conceptual modelling |
| UML | Unified modelling language |
| SysML | Systems modelling language |
| BSO | Bunge systemist ontology |
| IS | Information systems |
| BWW | Bunge, wand & weber ontology |
| CESM | Composition, environment, structure and mechanisms model |
| ER | Entity-relationship, it is a data model |
| CTT | Concurrent task |
| BPMN | Business process modelling notation |
| OLAP | Online analytical processing |
| CD | Class diagram, it is a UML diagram |
| PAD | Pathway analysis diagram |
| CSHG | Conceptual schema of the human genome |
| IPA | Ingenuity pathway analysis |
| EHR | Electronic health record |
| OntoUML | An ontologically well-founded language for ontology-driven CM; it is built as a UML extension based on UFO |
| UFO | Unified Foundational Ontology |
| ODCM | Ontology-driven conceptual modelling |

### Acknowledgements
This paper is an extended version of "The notion of "System" as a core conceptual modelling construct for the life sciences", presented at the International Conference on Conceptual Modelling (ER 2021), Conceptual Modelling for Life Sciences, and published in the proceedings by Springer Nature. https://doi.org/10.1007/978-3-030-89022-3_28.

### About this Supplement
This article has been published as part of *BMC Bioinformatics Volume 23 Supplement 11, 2022: Selected articles on Conceptual Modeling for Life Sciences (CMLS 2021 workshop and ER 2021 conference)*. The full contents of the supplement are available online at https://bmcbioinformatics.biomedcentral.com/articles/supplements/volume-23-supplement-11.

### Author contributions
RL has studied in depth Bunge's works in order to propose the systemist extension that has been conceptually characterized and presented in this work for the Life Science domain. VS has explored concrete life sciences working contexts of application and she has been a major contributor in writing the manuscript. OP has developed the notation that has been applied to the selected working domain examples. All authors participated to the discussions and read and approved the final manuscript.

### Funding
This work has been developed with the financial support of the Generalitat Valenciana and the Valencian Innovation Agency under the projects PROMETEO/2018/176 and INNEST/2021/57 and co-financed with ERDF. It was also supported by the J. Mack Robinson College of Business, Georgia State University.

### Availability of data and materials
Not applicable.

## Declarations

### Ethics approval and consent to participate
Not applicable.

### Consent for publication
Not applicable.

### Competing interests
The authors declare that they have no competing interests.

## Publisher's Note